\documentclass[aps,twocolumn,showpacs,groupedaddress]{revtex4}
\usepackage{amssymb,amsmath}
\usepackage{graphicx,array}
\usepackage{dcolumn}
\usepackage{bm}
\usepackage[normalem]{ulem}
\usepackage[usenames,dvipsnames]{color}

\begin{document}
\title{Electrostatic engineering of strained ferroelectric perovskites from first-principles}

\author{Claudio Cazorla}
\affiliation{School of Materials Science and Engineering,
             UNSW Australia, Sydney, NSW 2052, Australia}
\affiliation{Integrated Materials Design Centre, UNSW Australia,
             Sydney, NSW 2052, Australia}

\author{Massimiliano Stengel}
\affiliation{Institut de Ci$\grave{e}$ncia de Materials de Barcelona
            (ICMAB-CSIC), 08193 Bellaterra, Spain}
\affiliation{ICREA - Instituci\'o Catalana de Recerca i Estudis Avan\c{c}ats, 08010 Barcelona, Spain}

\begin{abstract}

Design of novel artificial materials based on ferroelectric perovskites relies on    
the basic principles of electrostatic coupling and in-plane lattice matching.
These rules state that the out-of-plane component of the electric displacement field
and the in-plane components of the strain are preserved across a layered                  
superlattice, provided that certain growth conditions are respected. Intense
research is currently directed at optimizing materials functionalities based
on these guidelines, often with remarkable success.
Such principles, however, are of limited practical use unless one disposes of 
reliable data on how a given material behaves under arbitrary electrical
and mechanical boundary conditions. 
Here we demonstrate, by focusing on the 
prototypical ferroelectrics PbTiO$_3$ and BiFeO$_3$ as testcases, how such information 
can be calculated from first principles in a systematic and efficient way.     
In particular, we construct a series of two-dimensional maps that describe the 
behavior of either compound (e.g. concerning the ferroelectric polarization and  
antiferrodistortive instabilities) at any conceivable choice of the in-plane
lattice parameter, $a$, and out-of-plane electric displacement, $D$.
In addition to being of immediate practical applicability to superlattice design,
our results bring new insight into the complex interplay of competing degrees of 
freedom in perovskite materials, and reveal some notable instances where the behavior
of these materials depart from what naively is expected.

\end{abstract}

\maketitle

\section{Introduction}
\label{sec:intro}
\begin{figure}
\centerline{
\includegraphics[width=1.0\linewidth]{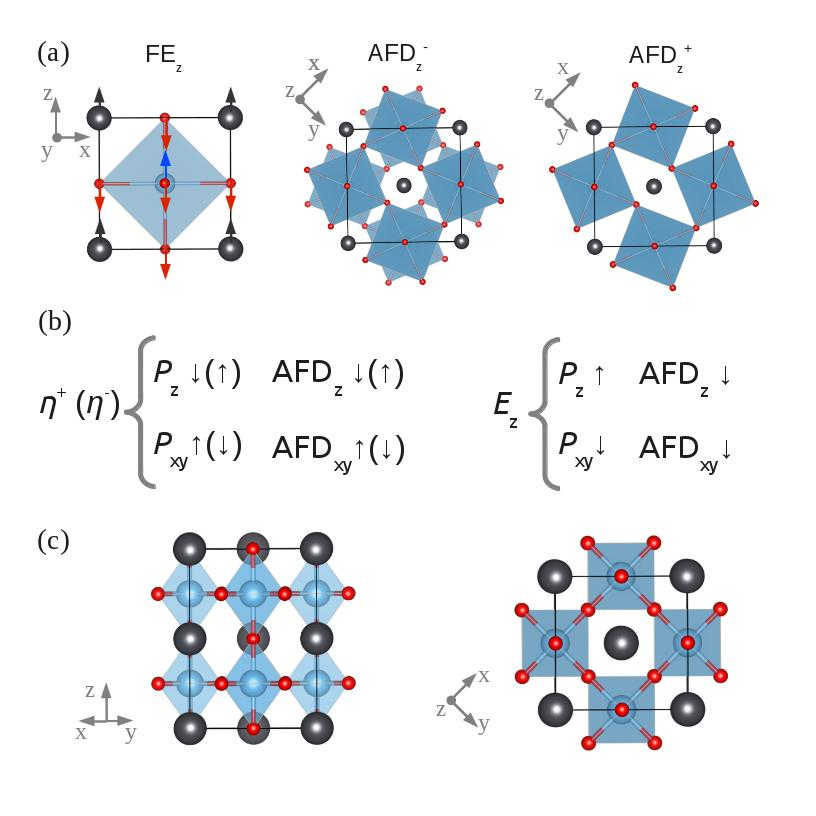}}
\vspace{-0.40cm}
\caption{(a)~Representation of typical FE and AFD distortions in pervoskite oxides.
         (b)~Expected trends of the polarization and octahedral O$_{6}$ rotations 
             in oxide films upon external mechanical and electric field perturbations. 
             $\eta^{+(-)}$ represents a tensile~(compressive) in-plane strain, $E_{z}$ 
             an out-of-plane electric field, and $\uparrow~(\downarrow)$ an 
             increase~(decrease) in the corresponding quantity.  
         (c)~Sketch of the $20$-atom $\sqrt{2} \times \sqrt{2} \times 2$ simulation cell  
         used in our calculations; red, blue and black spheres represent O, $B$, 
         $A$ atoms in $AB{\rm O}_{3}$ perovskites, respectively. The corresponding 
         lattice vectors are ${\bf a_{1}} = (a, a, 0)$, ${\bf a_{2}} = (a, -a, 0)$, 
         and ${\bf a_{3}} = (0, 0, 2a)$, where $a$ is the lattice parameter of the
         $5$-atom primitive perovskite cell.}
\label{fig-intro}
\end{figure}

Perovskite-structure oxides with ABO$_{3}$ formula display a wide range
of functional properties that are sought after for applications in energy
and nanoelectronics.
Such functionalities are, to a large extent, governed by a number of 
symmetry-lowering lattice instabilities, which naturally occur in the crystals 
depending on the chemical nature of the A and B cations.
Ferroelectric (FE) and antiferrodistortive (AFD) modes are by far the 
most common distortions in ABO$_{3}$ compounds. 
The former are typically associated with ``soft'' (i.e. unstable) zone-center phonons of 
the reference cubic phase (see Fig.~\ref{fig-intro}a)~\cite{cohen90,cohen92}.
Whenever they dominate, the crystal possesses a spontaneous (and switchable) 
electric polarization, which can be exploited for information storage and 
(via the piezoelectric effect) for electromechanical transduction. 
The latter, on the other hand, are related to zone-boundary instabilities, whereby 
the oxygen octahedral cages surrounding the B cation cooperatively rotate (either 
in antiphase or in phase) about a particular axis (see Fig.~\ref{fig-intro}a). 
While AFD modes are non-polar, they are extraordinarily important for
the physics of magnetic perovskites, where the couplings between 
neighboring transition-metal sites are sensitive to the B$-$O$-$B 
bond angle. 
Cases where FE polarization and AFD distortions are present and mutually coupled 
in the same phase are especially intriguing, both from a practical and a fundamental 
perspective. 
Indeed, recent research has unvealed a number of notable examples where 
octahedral rotations can strongly enhance, or even induce, a ferroelectric distortion
in materials that would be otherwise nonpolar.~\cite{junquera12,bousquet08}
Most importantly, such unconventional forms of ferroelectricity, often referred 
to as ``improper''~\cite{levanyuk74} or ``hybrid improper''~\cite{benedek11,benedek12}, 
appear as a promising route for achieving electrical control of magnetism in multiferroics 
and magnetoelectrics, a tantalizing goal that has remained so far elusive in spite of the 
numerous recent breakthroughs~\cite{heron14,fusil14}.

Practical ways to manipulate and design the aforementioned functionalities 
are currently being investigated along two main avenues. These consist in
either synthesizing new perovskite materials (e.g. by playing with the A- 
and B-site chemistry), or in altering the behavior of existing ones via external 
perturbations. 
In the context of the latter strategy, a large part of the recent efforts 
have been directed at appropriately modifying the strength of the FE and AFD 
instabilities (and/or their mutual interplay) via the imposed 
mechanical~\cite{hatt10,rondinelli11,rondinelli11b,may10}
and electrostatic boundary conditions~\cite{wu12,stengel12b,hong13} 
(see Fig.~\ref{fig-intro}b for a summary of the most common trends). 

\emph{Strain engineering}, for instance, consists in growing the material in 
a thin-film form, where bonding to the substrate favors coherency in the in-plane 
registry.
Thanks to the availability of numerous substrates with a variety of lattice
parameters, such a practice has become very popular in the past few years~\cite{dawber05b,choi04}.
Polar (FE) and structural (AFD) degrees of freedom are generally sensitive  
to the applied strain, and this often allows for the stabilization of phases
with enhanced functional properties which would be otherwise inaccessible at 
the bulk level~\cite{bea09,warusawithana09}.
(The regions of parameter space that lie close to phase boundaries are 
particularly responsive to external perturbations because of their structural
``softness''~\cite{jacek10}.)

Control of electrostatics in perovskites, on the other hand, is comparatively more subtle,
as it can be easily thwarted by a redistribution of the mobile charges (e.g. 
electron/hole carriers, ionic defects, etc.) or the formation of ferroelectric 
domains. Also, leakage currents often limit the magnitude of the electric field that
can be realistically applied via an external voltage source~\cite{junquera11}.
Yet, the so-called \emph{electrostatic coupling}~\cite{junquera11,ghosez08,zubko12,dawber12,dawber05,dawber07} 
is one of the most valuable design principles in ferroelectric superlattices based on perovskites.
It states that, at the boundary between two insulating layers, the 
normal component of the electric displacement field must be preserved;
this effectively constrains the out-of-plane polarization state of the 
stack to a common value that, depending on the materials choice and
conditions, might be markedly different than the spontaneous polarization
of the parent compounds.
In the vast majority of cases, electrostatics and strain work together, 
and need to be simultaneously taken into account when interpreting the
experiments with quantitatively predictive models.

First-principles simulations within the framework of density-functional theory
(DFT) have been of invaluable help in rationalizing and guiding the experimental
efforts towards synthesis of new materials with enhanced properties.
Until few years ago DFT methods were restricted to zero-electric-field 
simulations, which complicated the description of electrostatics 
in multicomponent oxide systems. Such effects have traditionally been 
described by means of phenomenological models, where the DFT input 
was limited to the computation of the relevant coupling coefficients. 
Recent and timely advances in first-principles computational methods have now 
enabled full control over the macroscopic electrical variable of interest  
(i.e., electric field, polarization and electric displacement) in the simulation of 
ferroelectric, magnetoelectric, and piezoelectric materials~\cite{souza02,dieguez06,stengel09c}. 
These methods now allow to systematically and accurately explore both strain and 
electrostatic effects at the level of the individual bulk material. 
In particular, by calculating the ground-state of a given 
compound as a function of the in-plane strain and out-of-plane electric displacement,
one can trace two-dimensional maps that comprehensively describe \emph{all  
configurations} that are, in principle, accessible in a superlattice geometry, i.e. at arbitrary 
electrical and mechanical boundary conditions.

Among the variety of materials that have been considered as
building blocks for ferroelectric superlattices, perovskite-structure 
PbTiO$_{3}$ (PTO) and BiFeO$_{3}$ (BFO) occupy a special place.
Indeed, the former is the end member of the Pb$_x$Zr$_{1-x}$TiO$_3$ family,
the technologically most important piezoelectric material, while the latter
is the most intensely studied of all naturally occurring multiferroics.
Interestingly, even if both materials present active (and strong) FE and AFD 
instabilities in their cubic reference phase, their ground state is remarkably different.
On one hand, PTO condenses in a purely ferroelectric $P4mm$ phase, where 
the AFD instability has been suppressed by the large polar distortion. (FE 
and AFD modes generally, although not always~\cite{benedek11,benedek12}, compete.)
Note that the AFD modes, even if they are absent in the low-temperature
bulk phase, have been shown to play an important role in the ferroelectric~$\to$~paraelectric 
phase transition~\cite{wojdel13}, and are predicted to become active in thin films at 
medium to large values of the epitaxial strain~\cite{yang12}.
Bulk BFO, on the other hand, adopts a rhombohedral ($R3c$) 
structure where a large polarization and antiphase AFD distortions 
[both oriented along the pseudocubic (111) direction] coexist.
The balance between polar and non-polar instabilities is extraordinarily subtle in this 
latter material, leading to a rich and complex energy landscape with many local 
minima~\cite{dieguez11}. 
In fact, numerous phase transitions driven by 
pressure~\cite{guennou11}, temperature~\cite{prosandeev12,rahmedov12,cazorla13}, and expitaxial 
strain~\cite{bea09,zeches09} have been recently predicted and observed in BFO.
Given the extraordinarily rich behavior of both materials, it is hardly 
surprising that their combination results in an even broader variety of original 
phenomena. Indeed, recent \emph{ab initio} simulations have shown
that epitaxial short-period superlattices made of PbTiO$_{3}$ and BiFeO$_{3}$ 
layers may exhibit strain-driven isostructural phase transitions, exotic 
metastable phases, and unusually large oxygen octahedra tilts~\cite{yang12,stengel12,cazorla14}.
In view of the great fundamental and technological interest of bulk PTO, BFO, and  
layered heterostructures comprising both species, a detailed knowledge of how  
mutually coupled FE and AFD degrees of freedom in these materials 
respond to external electric fields and/or mechanical constraints, is highly 
desirable. 

In this article, we present a systematic first-principles study of the energy, 
structural and dielectric properties of ``bulk''~\footnote{We perform
bulk calculations, but bearing in mind a thin-film and/or superlattice 
geometry, where such conditions of strain and electric field can be 
experimentally achieved.}
PTO and BFO as a function of \emph{both} in-plane epitaxial strain ($a$) and 
out-of-plane electric displacement field ($D$). 
Our computational results are presented in the form of two-dimensional $(a, D)$ maps, 
which describe the structural and electronic ground state of the material at
an arbitrary choice of the electrical and mechanical boundary conditions.
These maps provide a comprehensive overview of \emph{all} the phases of 
BFO and PTO that can be realized via strain engineering and electrostatic
coupling, i.e. by growing the material in a thin-film or superlattice 
form. Most importantly, these maps illustrate how the main functional 
properties (e.g. dielectric, piezoelectric) of either material evolve
as a function of $(a,D)$, and thus constitute an invaluable guide 
in the experimental search for new artificial materials.
For instance, we identify the specific regions of the phase
diagram where PTO or BTO display an unusual dielectric and structural 
softness, leading to very large dielectric and piezoelectric
responses. For PTO, this region is located at small positive strain,
where the polarization abruptly rotates from out-of-plane to in-plane.
(Interestingly, in a very small range of in-plane strain values,
we find a stable monoclinic structure, where the polarization
has both in-plane and out-of-plane components and is accompanied
by an out-of-phase AFD tilt mode.)
In the case of BFO, such a structural softness occurs in the 
orthorhombic $Pmc2_{1}$ phase that becomes stable at high tensile
strain, and whose existence was recently predicted by means of 
first-principles calculations~\cite{yang12,stengel12}.

In addition to their practical interest, our results also provide a wealth of 
valuable insight into the physics of either material, especially regarding 
the mutual interplay of strain, polarization and AFD tilts.
In this context, parallel to rationalizing the results of earlier
works on PTO- and BFO-based systems~\cite{junquera12,yang12,stengel12} 
we could identify a notable case where the main order parameters
do not follow the typical behavior that has been traditionally assumed in 
phenomenological theories. 
In particular, both PTO and BFO are characterized, at high tensile strain,
by an in-phase out-of-plane AFD mode, whose amplitude \emph{grows}, rather 
than decreasing, when an out-of-plane electric field (of arbitrary strength) is applied.
This is in stark constrast with the expected trends (AFD modes usually tend
to compete with the polar degrees of freedom) represented in Fig.~\ref{fig-intro}b.

This work is organized as follows: In Sec.~\ref{sec:methods} we provide 
the details of the computational strategy and \emph{ab initio} methods that we used in the 
construction of the $(a, D)$ diagrams. In Sec.~\ref{sec:results}, we present our results 
and discuss them extensively. Finally, we conclude by summarizing our main findings 
in Sec.~\ref{sec:summary}.

\section{Computational Details}
\label{sec:methods}

\subsection{Construction of the $(a, D)$ maps}
\label{subsec:computational}
Our calculations are performed 
within the local spin density approximation to density-functional theory,
as implemented in the ``in-house'' LAUTREC code.
We apply a Hubbard $U = 3.8$~eV to the Fe ions~\cite{yang12,kornev07}.
In all the simulations, we use the 20-atom $\sqrt{2} \times \sqrt{2} \times 2$ 
simulation cell depicted in Fig.~\ref{fig-intro}c, which allows for reproducing 
the ferroelectric and antiferrodistortive distortions of interest (i.e., in-phase 
AFD$^{+}$ and out-of-phase AFD$^{-}$, see Fig.~\ref{fig-intro}a).

It is worth noting that working at fixed electric displacement $D$  
is in many respects analogous to working at fixed electric field $E$. 
At fixed $D$, however, a much broader range of 
configurations can be explored in FE materials, including regions of parameter
space where the system is unstable with respect to a polar distortion~\cite{stengel09c,stengel09}.
Controlling $D$ is akin to controlling $P$, and therefore provides an
intuitive and direct conceptual link to phenomenological descriptions
(i.e., Landau theory) where $P$ is treated as the main order parameter. 

Atomic and cell relaxations are performed by constraining the value of the 
out-of-plane component of $D$~\cite{stengel09c} and the two in-plane lattice vectors 
${\bf a_{1}}$ and ${\bf a_{2}}$ (see Fig.~\ref{fig-intro}). 
In the remainder of this article, $D$ will stand for the out-of-plane component 
of the electric displacement vector.
Small tolerances of $0.005$~eV/\AA~ and $0.001$~Kbar are imposed on the atomic forces 
and stresses. 
We correct for the bias deriving from the Pulay stress in the out-of-plane direction, 
$\pi_{\rm P}$, by introducing a compensating $-\pi_{\rm P} \Omega$ term (where $\Omega$ 
is the cell volume) in the expression of the internal energy. The calculations are repeated 
in order to span the physically relevant space defined by the set of parameters $(a,D)$.

In order to achieve an accurate description of the energy and structural properties 
of PTO and BFO in the $3.6 \le a \le 4.2$~\AA~ and $0 \le D \le 1.5~e/S$ intervals, 
we carry out explicit calculations on a fine regular grid of $(a,D)$ points, spaced by 
$\Delta a = 0.05$~\AA~ and $\Delta D = 0.1~e/S$ (where $e$ is the electron charge 
and $S$ the surface of the five-atom unit cell). Spline interpolations are 
subsequently applied to the calculated data points, in order to obtain smooth 
two-dimensional maps (see Figs.~\ref{figpto-1}-\ref{figbfo-2}).

\subsection{Electrostatic and distorsion analysis}
\label{subsec:distorsions}
The components of the total polarization in each relaxed configuration, 
i.e., in-plane $\mathbf{P}_{\parallel} = (P_{x}, P_{y})$ and out-of-plane 
$P_{z}$, were computed with the formula
\begin{equation} 
P_{\alpha} = \frac{1}{\Omega} \sum_{\kappa\beta}  Z_{\kappa\beta \alpha}^{*} u_{\kappa\beta}~, 
\label{eq:polarization}
\end{equation} 
where $\Omega$ is the volume of the cell, $\kappa$ runs over all the atoms there, 
$\alpha,\beta = x, y, z$ represent Cartesian directions, $\bf{u}_{\kappa}$ is the 
displacement vector of the $\kappa$-th atom as referred to the cubic perovskite 
phase, and $\boldsymbol{Z}^{*}_{\kappa}$ the Born effective charge tensor calculated 
in the paraelectric reference state.  
We note that since both PTO and BFO materials are good ferroelectrics, one can regard 
the out-of-plane component of the total polarization, $P_{z}$, to be approximately 
equal to $D$.

The magnitude of the antiferrodistortive AFD$_{z}^{+}$, AFD$_{z}^{-}$ (both with
the rotation axis oriented along $[001]$) and AFD$_{xy}^{-}$ 
(with $[110]$ axis) rotations were evaluated by considering the projection 
of the $\bf{u}_{\kappa}$ vectors on the corresponding zone-boundary eigenmode, 
see Fig~\ref{fig-intro}a. (We use a ``$+$'' or a ``$-$'' symbol to indicate
in-phase and out-of-phase tilt modes, respectively.)

\subsection{Computation of functional properties}
\label{subsec:funct-prop}
The availability of the full $(a,D)$ maps for a given material
allows us to extract important functional properties (e.g. dielectric 
and piezoelectric coefficients) in a straightforward way as a by-product of our 
calculations~\cite{stengel09c}.
In particular, the inverse dielectric constant of the system, $\epsilon^{-1}$ 
(by $\epsilon$ without subscripts we
shall always imply the 33 component of the dielectric tensor),
adopts the simple form
\begin{equation}
\epsilon^{-1} = \frac{4 \pi}{\Omega} \frac{{\rm d}^{2} U}{{\rm d} D^{2}}~,
\label{eq:dielectric}
\end{equation}
where $U$ is the internal energy, $\Omega = S c$ the unit cell volume, $S$ the unit cell surface, 
and $c$ the unit cell out-of-plane parameter. 
Also, the \emph{longitudinal} piezoelectric coefficient is readily obtained as
\begin{equation}
d_{33} = \frac{dc}{dV} = \left( \frac{dV}{dD} \right)^{-1} \frac{dc}{dD} = \frac{\epsilon}{c} \frac{dc}{dD}~, 
\label{eq:d33}
\end{equation}
where $V =\frac{4 \pi}{S} \frac{dU}{dD}$ is the potential drop across the unit cell in the $z$ 
direction~[\onlinecite{stengel09c}].
Likewise, the \emph{shear} piezoelectric coefficient is expressed within our formalism 
as
\begin{equation}
d_{35} = \frac{dw}{dV} = \left( \frac{dV}{dD} \right)^{-1} \frac{dw}{dD} = \frac{\epsilon}{c} \frac{dw}{dD}~, 
\label{eq:d15}
\end{equation}
where $w$ corresponds to the projection of the ${\bf a_{3}}$ lattice vector on the basal plane 
defined by ${\bf a_{1}}$ and ${\bf a_{2}}$ (see Fig.~\ref{fig-intro}).
(The Voigt notation of the shear strain, therefore, refers to the rotated axes of the
20-atom cell, and not to the pseudocubic axes.)

The response coefficients defined in Eq.~(\ref{eq:dielectric}), Eq.~(\ref{eq:d33}) and Eq.~(\ref{eq:d15})
are directly related to those discussed by Wu, Vanderbilt and Hamann (WVM)~\cite{wu05}. In particular, the
dielectric constant of Eq.~(\ref{eq:dielectric}) corresponds to the free-stress permittivity of WVM, 
$\epsilon_{33}^{(\sigma)}$. Regarding the $d_{\alpha \beta}$ tensor, our notation is consistent with WVM. 
It is useful for later use (again, following WVM) to introduce the fixed-$D$, free-stress piezoelectric coefficients
\begin{equation}
g_{33} = \frac{1}{c} \frac{dc}{dD}, \qquad g_{35} = \frac{1}{c} \frac{dw}{dD},
\end{equation} 
which trivially relate to the corresponding component of the ${\bf d}$-tensor as
$d_{\alpha \beta} = \epsilon g_{\alpha \beta}$.

Note that the above definitions can be readily used to predict the dielectric 
and piezoelectric properties of a multicomponent system (e.g. a layered superlattice) 
based on the bulk properties of the constituents.
For example, consider a superlattice (SL) with a total thickness
$t_{\rm A}$ of material A and $t_{\rm B}$ of material B. In this case, one has
\begin{eqnarray}
\frac{t_{\rm A} + t_{\rm B}}{\epsilon^{\rm SL}} &=& \frac{t_{\rm A}}{\epsilon^{\rm A}} +
  \frac{t_{\rm B}}{\epsilon^{\rm B}}, \nonumber \\
(t_{\rm A} + t_{\rm B})g^{\rm SL} &=& t_{\rm A} g^{\rm A} +
  t_{\rm B} g^{\rm B}~. 
\label{eq:gadd}
\end{eqnarray}
As above, one can then easily recover the overall $d$-coefficients as
$d^{\rm SL} = \epsilon^{\rm SL} g^{\rm SL}$.

\section{Results and Discussion}
\label{sec:results}

\subsection{Bulk PbTiO$_{3}$}
\label{subsec:pto}

\begin{figure}
\includegraphics[width=1.00\linewidth]{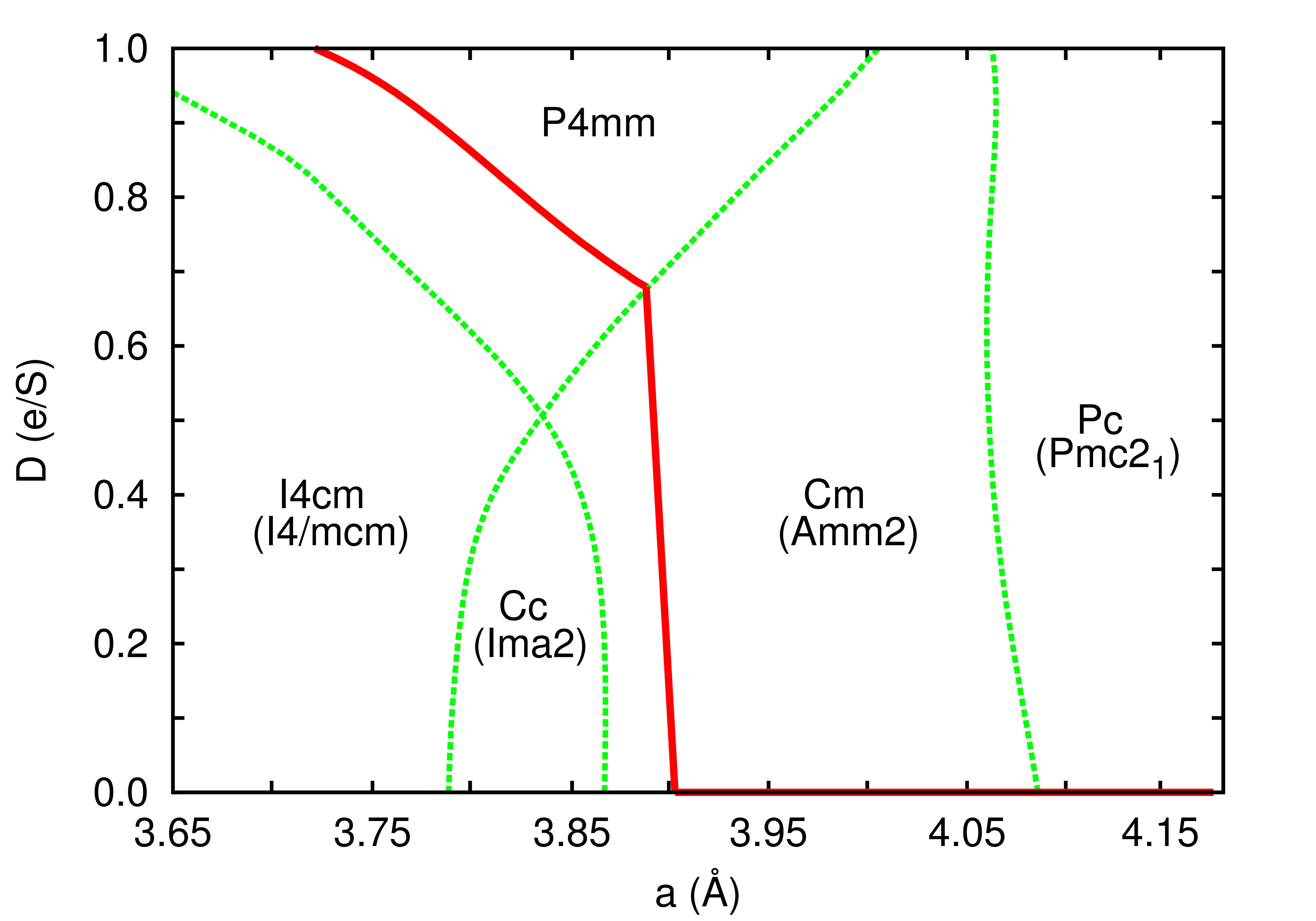}
\caption{Calculated phase diagram of PbTiO$_{3}$ as a function 
         of the in-plane lattice parameter and out-of-plane electric 
         displacement. The thick red line joins the states of minimum 
         energy obtained at fixed $a$, and the dashed green lines 
         represent two-phase boundaries involving second-order phase 
         transitions (see text). Space groups quoted within 
	 parentheses correspond to crystal structures found at $D = 0$.} 
\label{pto-pd}
\end{figure}

\begin{figure*}
\includegraphics[width=1.00\linewidth]{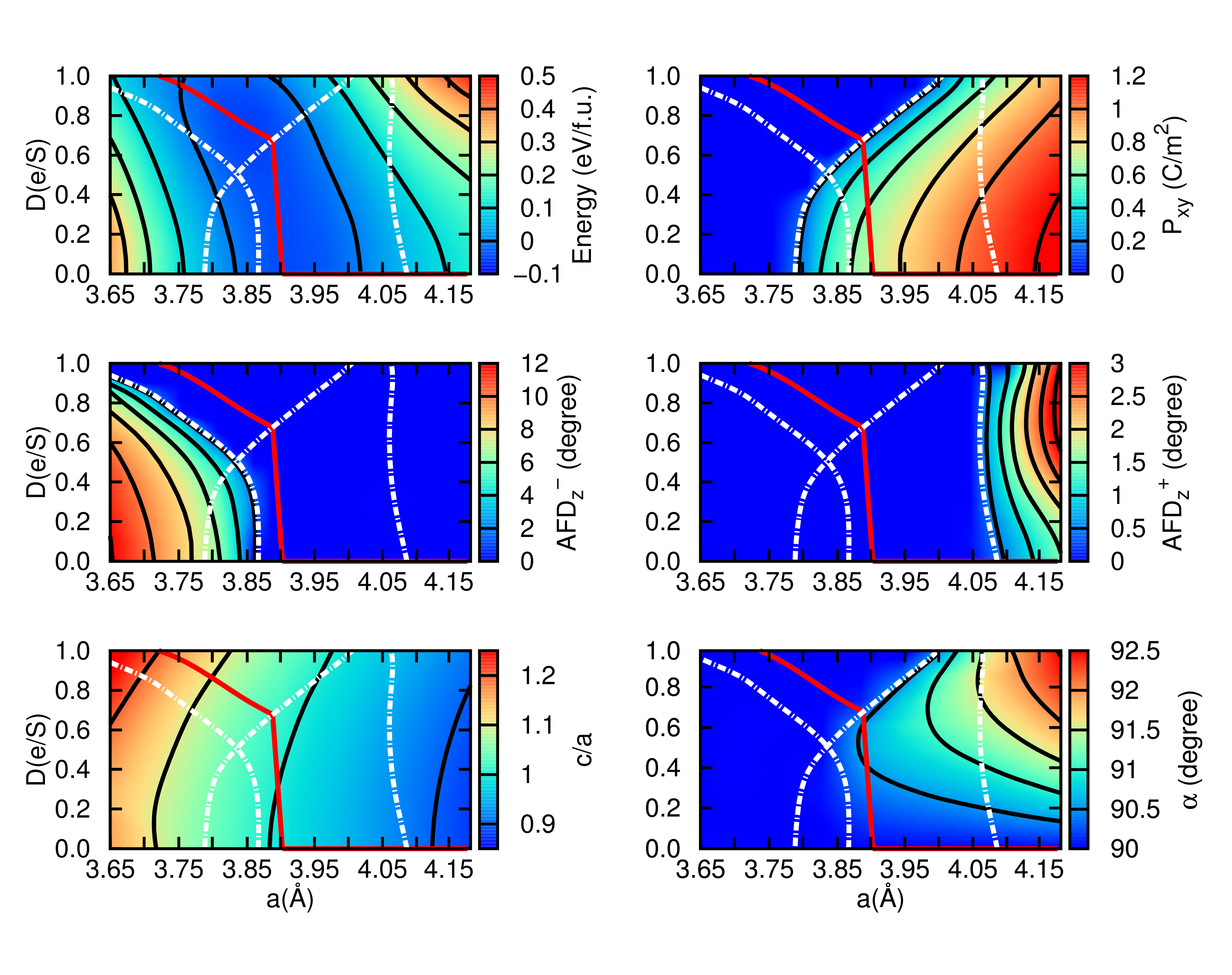}
\vspace{-1.25cm}
\caption{Calculated energy and structural properties of PTO expressed as a function 
         of $a$ and $D$. $\alpha$ is the monoclinic angle, i.e. that formed by the ${\bf a_{3}}$ 
         and ${\bf a_{1}} + {\bf a_{2}}$ vectors in the relaxed unit cell; 
         $P_{xy}$ the in-plane polarization (oriented along the $[110]$ direction); 
         $c/a$ is the axial ratio $|{\bf a_{3}}|/(\sqrt{2}a)$. 
         Solid black curves are isolines connecting the states that share a common
         value for a given property, and the thick red curves joins the 
         states at electrostatic (but not necessarily mechanical) equilibrium.
         The dashed white lines match the dashed green curves of Fig.~\ref{pto-pd}, and
         indicate second-order phase boundaries.}
\label{figpto-1}
\end{figure*}

In Fig.~\ref{pto-pd}, we show a schematic phase diagram of PTO, calculated as a  
function of $(a, D)$, {while in Fig.~\ref{figpto-1} we report a series of
detailed 2D maps that illustrate the evolution of most relevant structural and 
electrical degrees of freedom within the same $(a,D)$ range.
For each region of the phase diagram (Fig.~\ref{pto-pd}) we indicate the space group of the 
stable crystal phase therein and we mark the resulting two-phase boundaries as a dashed line. 
(The same phase boundaries are also shown in the maps of Fig.~\ref{figpto-1}.)
Moving from smaller to larger values of $a$, the dashed lines correspond, respectively, 
to the suppression of AFD$_{z}^{-}$ tilts, the onset of $P_{xy}$ and the onset of 
AFD$_{z}^{+}$ rotations (see Fig.~\ref{figpto-1}).  
We note that, according to our calculations, all the phase transitions occurring in PTO are
of second-order type, i.e., the energy of the crystal varies continuously while specific 
distortions -- either FE, AFD, or both simultaneously -- emerge or are suppressed.
The solid red line in Figs.~\ref{pto-pd} and~\ref{figpto-1} indicates, 
for each value of $a$, the value of $D$ that minimizes the energy;
the points of this curve therefore correspond to the electrostatic equilibrium 
configurations as a function of the in-plane strain.
Note that $P_z$ exactly coincides with $D$ along such path, as at the
stationary points of the $U(D)$ curve the macroscopic electric field 
vanishes by definition.
The curve mostly follows the known trends for PTO thin films:~\cite{yang12,stengel12}
(i)~A zero or negative in-plane strain favors a tetragonal $P4mm$ state with a 
strong out-of-plane component of the polarization, $P_z$. PTO remains in the 
tetragonal $P4mm$ phase down to $a = 3.60$~\AA, where the resulting $c/a$ ratio 
and out-of-plane polarization become very large ($1.35$ and $1.36$~C/m$^{2}$, respectively).
(ii)~$P_z$ is suppressed upon application of a tensile strain, whereby the polarization 
aligns with the in-plane $(110)$ direction (the orientation of the polarization abruptly changes 
from $[110]$ to $[001]$ at $a = 3.90$~\AA) and the system transitions to the orthorhombic 
$Amm2$ phase. This is analogous to the phase occurring in BaTiO$_{3}$ under similar conditions~\cite{dieguez04}.
Finally, (iii)~at a large enough value of the tensile strain a second phase transition occurs,  where an
AFD$^+_z$ mode appears; as a consequence, the symmetry of the system reduces to $Pmc2_1$ as recently 
reported in Refs.~\cite{yang12,stengel12}.

While the behavior described above is fully consistent with the results of earlier studies we found, however,
some interesting surprises as well. First, we find a stable monoclinic $Cm$ phase in the 
narrow $3.89 \le a \le 3.91$~\AA~ interval. This phase is characterized by $P_{z},~P_{xy} \neq 0$ 
and $\alpha \neq 90^{\circ}$ (see Fig.~\ref{pto-cm}) and, to our knowledge, has never been 
predicted or observed previously. 
(The energy difference between the monoclinic $Cm$ and orthorhombic 
$Amm2$ phases is very small, i.e., $1-2$~mev/f.u., which explains why the
monoclinic structure might have been missed in earlier studies.)
Second, our calculated phase diagram of Fig.~\ref{pto-pd} clearly shows 
that AFD$_{xy}^{-}$ distortions do not occur in PTO at any value of 
$a$ or $D$.
Earlier literature studies~\cite{yang12,stengel12,blok11} predicted
such distortions to occur in the tensile-strain region, where the polarization
vector has an in-plane orientation. (These structures were, therefore, identified
as orthorhombic $Ima2$, rather than $Amm2$.)
This discrepancy can be rationalized by observing that the AFD$_{xy}^{-}$ mode
amplitude was found~\cite{yang12,stengel12,blok11} to be very small 
(i.e., $\sim 1^{\circ}$), which implies that that the $Amm2$ and $Ima2$ phases 
are essentially degenerate in energy. Whether they show up or not in a calculation
might therefore depend on subtle details of the numerical implementation.

Now, we shall discuss more in detail the results that we obtained by exploiting the 
full capabilities of our computational approach, i.e., by imposing arbitrary
$(a,D)$, and including those configurations that are characterized by a nonzero 
internal field along the $z$ direction.

\begin{figure}
\includegraphics[width=1.00\linewidth]{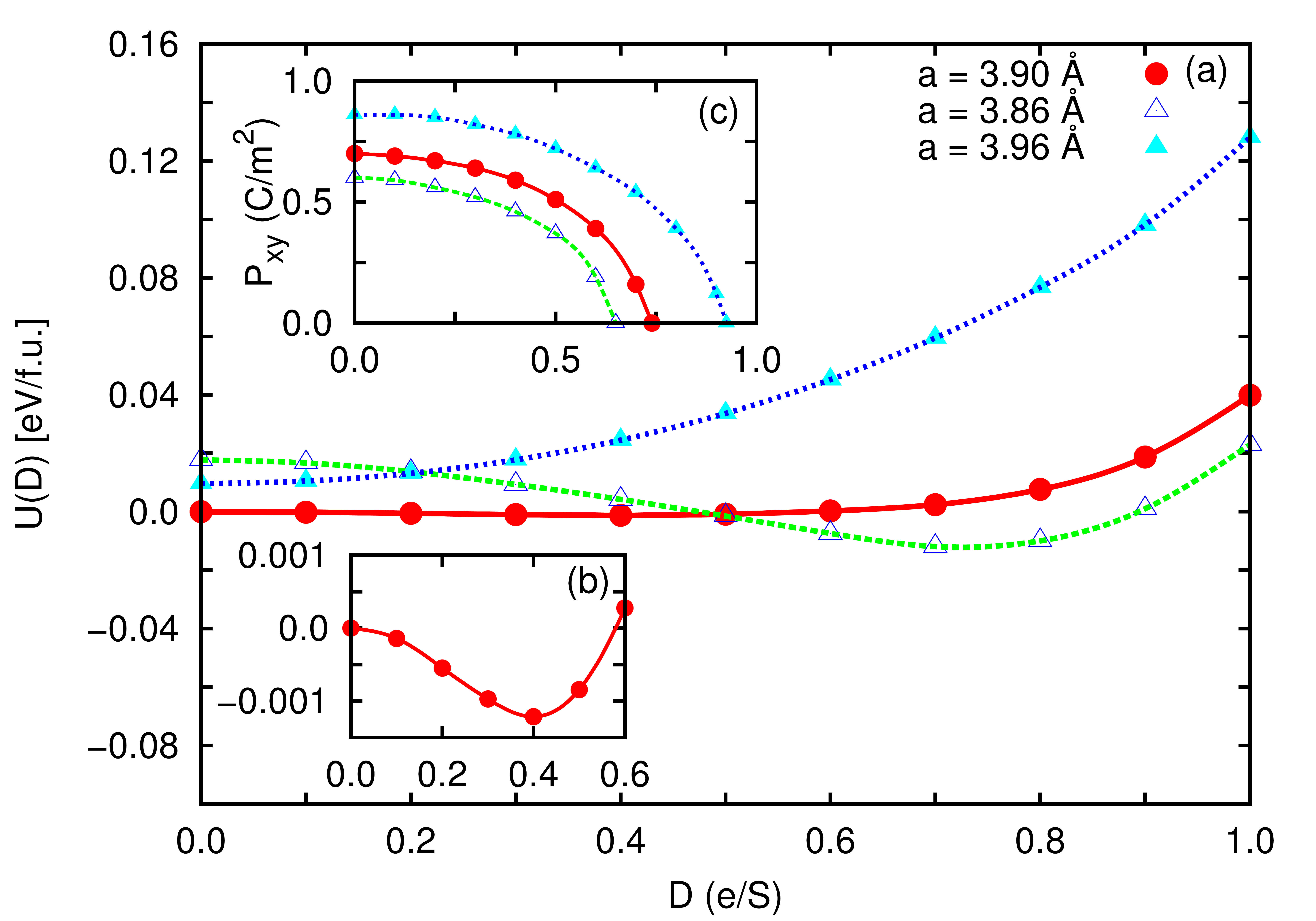}
\caption{(a)~Calculated energy curves in PTO at three different in-plane parameters 
          and expressed as a function of the electric displacement. (b)~Blow-up 
          of the energy function computed in the monoclinic $Cm$ phase. 
          (c)~In-plane polarization expressed as a function of $D$ at the three 
          selected in-plane strain values.}
\label{pto-cm}
\end{figure}

First of all, it is interesting to note the unusually flat potential-energy 
landscape in a vicinity of the zero-strain equilibrium state (see Fig.~\ref{figpto-1});
this indicates that the system is extremely sensitive to external perturbations in 
this region of the phase diagram.
Indeed, recall that in the orthorhombic $Amm2$~$\to$~tetragonal $P4mm$ 
transition the polarization abruptly rotates from in-plane to out-of-plane; 
thus, large dielectric and piezoelectric responses are expected to occur 
near $a = 3.90$~\AA  (we will comment more extensively on this point in 
Sec.~\ref{subsec:funct}).

The structurally most complex region of the phase diagram occurs in the moderately
compressive strain regime ($3.79 \le a \le 3.87$~\AA) and $D$ values that are 
smaller than the equilibrium $P_z$. 
Here all the most important structural modes of PTO ($P_{xy}$, $P_z$ and AFD$_{z}^{-}$ 
distortions) coexist, and the symmetry of the crystal is monoclinic $Cc$ (see 
Figs.~\ref{pto-pd} and~\ref{figpto-1}). (It reduces to orthorhombic $Ima2$ at $D = 0$.)
The occurrence of the $P_{xy}$ and AFD$_{z}^{-}$ distortions, which are active 
instabilities of the reference $P4/mmm$ structure but are absent in the zero-field
equilibrium phase within this strain interval, may appear surprising. This, 
however, can be easily rationalized in terms of the
mutual competition between different instabilities in PTO. 
At zero electric field (i.e., red curves in Figs.~\ref{pto-pd} and~\ref{figpto-1}), 
$P_z$ is large enough to prevail over the other degrees of freedom, which are completely 
suppressed. 
By reducing $P_z$ (as can occur, in practical situations, as a consequence of 
depolarizing field effects), the aforementioned cross-couplings are weakened, allowing
the competing $P_{xy}$ and AFD$_{z}^{-}$ distortions to reenter the ground-state
structure.
This finding explains the results of a recent study~\cite{junquera12}, where a superlattice 
geometry was found to induce a monoclinic structure analogous to that described here in the 
PTO layers.
Our results indicate that the electrical boundary conditions are, in fact, 
likely to be the main cause for such an outcome. 
Higher in-plane compression ($a \le 3.79$~\AA) suppress the in-plane polarization instabilities 
completely, and at the same time increase the strength of AFD$_{z}^{-}$ rotations; the 
symmetry  of the crystal is tetragonal $I4cm$ (tetragonal $I4/mcm$ at $D = 0$)
Our data suggest that the AFD$_{z}^{-}$ rotations will consistently show up at arbitrary 
(moderate to strong) compressive strain values, provided that a depolarizing effect is present.

In the moderate tensile strain regime ($3.90 \le a \le 4.09$~\AA), all oxygen octahedral 
rotations are suppressed and the symmetry of the system is monoclinic $Cm$. (This corresponds to
a polarization oriented along $[uuv]$, with $v<u$, where the out-of-plane component can be 
induced by applying an out-of-plane electric field to the equilibrium $Amm2$ phase.)
Finally, in the $4.09 \le a \le 4.20$~\AA~ interval, the monoclinic angle 
$\alpha$ increasingly  deviates from $90^{\circ}$ as $D$ is increased, 
and the AFD$_{z}^{+}$ rotations become more prominent. 
Note that, as $P_{z}$ increases, the in-plane component of the polarization 
becomes smaller (although full suppression of $P_{xy}$ is not observed). 
These features are consistent with a monoclinic $Pc$ phase that is closely related 
to the equilibrium orthorhombic $Pmc2_{1}$ found at $D = 0$. We note that the 
trend observed in this region is quite surprising: as $P_{z}$, or equivalently $D$, 
increases the AFD$_{z}^{+}$ rotations become larger.
This contrasts with the typical trends shown in Fig.~\ref{fig-intro}. (We already
pointed out that such a trend was indeed followed by the AFD$_z^-$ / $P_{xy}$ modes 
in the compressive regime.)
Thus, a cooperative rather than a competitive coupling appears to be operating 
between the $P_z$ and AFD$_{z}^{+}$ degrees of freedom. This observation points to 
new interesting opportunities for tuning of the AFD$_{z}^{+}$ oxygen rotations 
via application of an external electric field.

\subsection{Bulk BiFeO$_{3}$}
\label{subsec:bfo}

\begin{figure}
\includegraphics[width=1.00\linewidth]{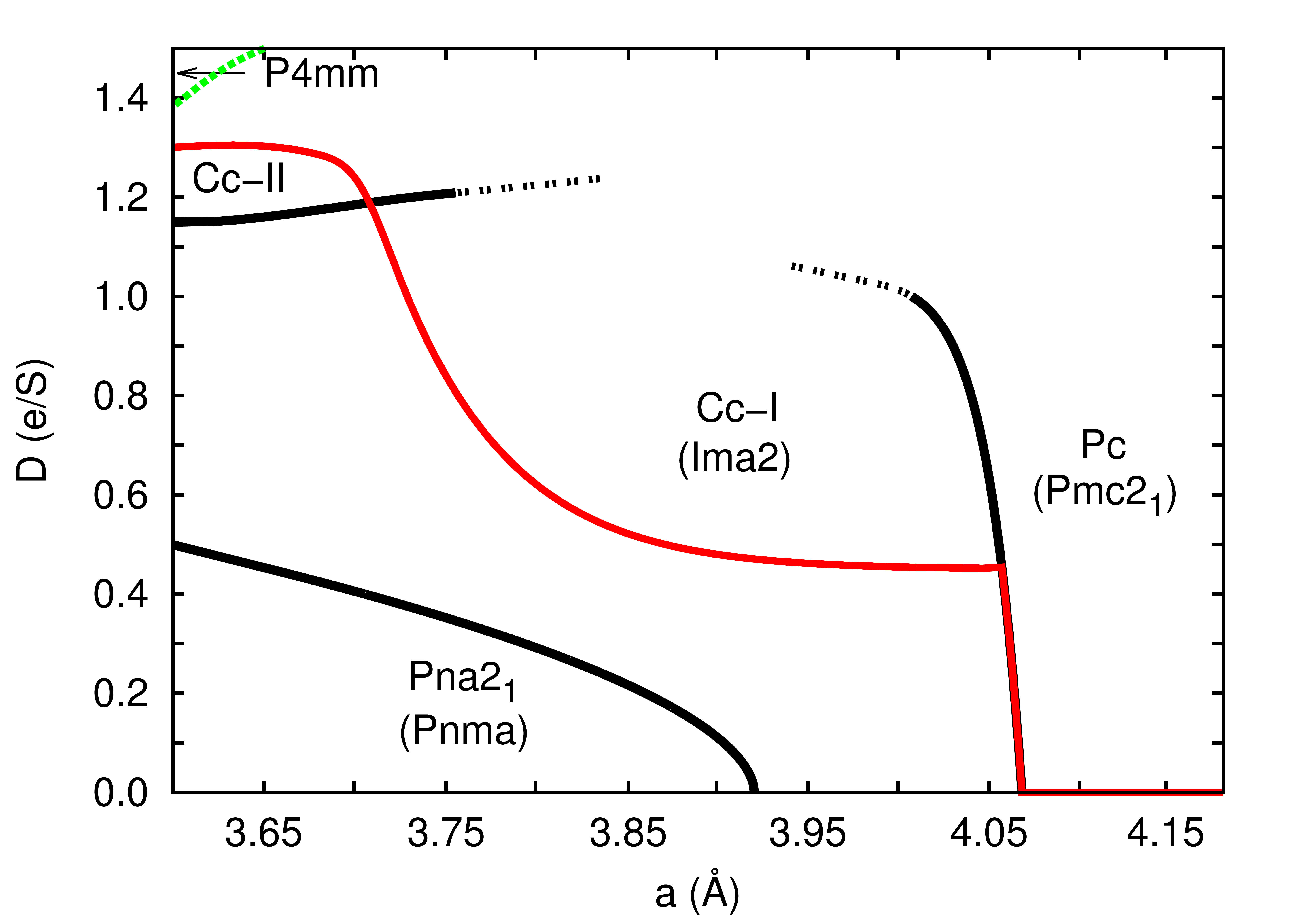}
\caption{Calculated phase diagram of BiFeO$_{3}$ as a function
         of the in-plane lattice parameter and out-of-plane electric
         displacement. The thick red line joins the states of minimum
         energy obtained at fixed $a$, and the solid black (dashed green) 
         lines represent two-phase boundaries involving first-order (second-order) 
         phase transitions. Space groups quoted within parentheses correspond
         to crystal structures obtained at $D = 0$. Dotted black lines indicate
         extrapolation.}
\label{bfo-pd}
\end{figure}

\begin{figure*}
\includegraphics[width=1.00\linewidth]{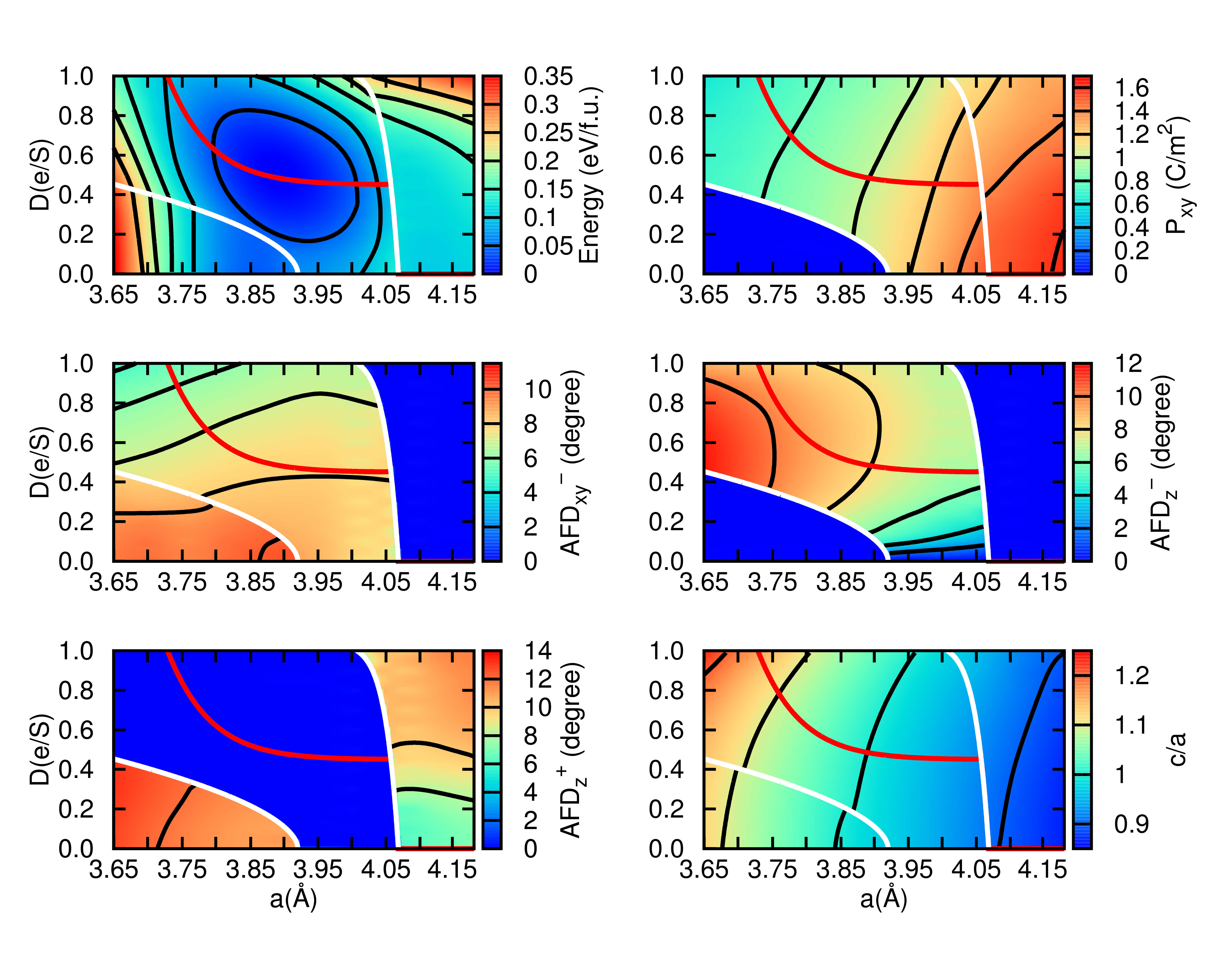}
\vspace{-1.25cm}
\caption{Calculated energy and structural properties of BFO expressed as a function 
         of $a$ and $D$. Solid black curves are isolines, and the thick red line 
         joins the states of minimum energy (zero electric field) at a given in-plane $a$. 
         Thick white curves indicate first-order phase transitions.}
\label{figbfo-1}
\end{figure*}

\begin{figure}
\includegraphics[width=1.00\linewidth]{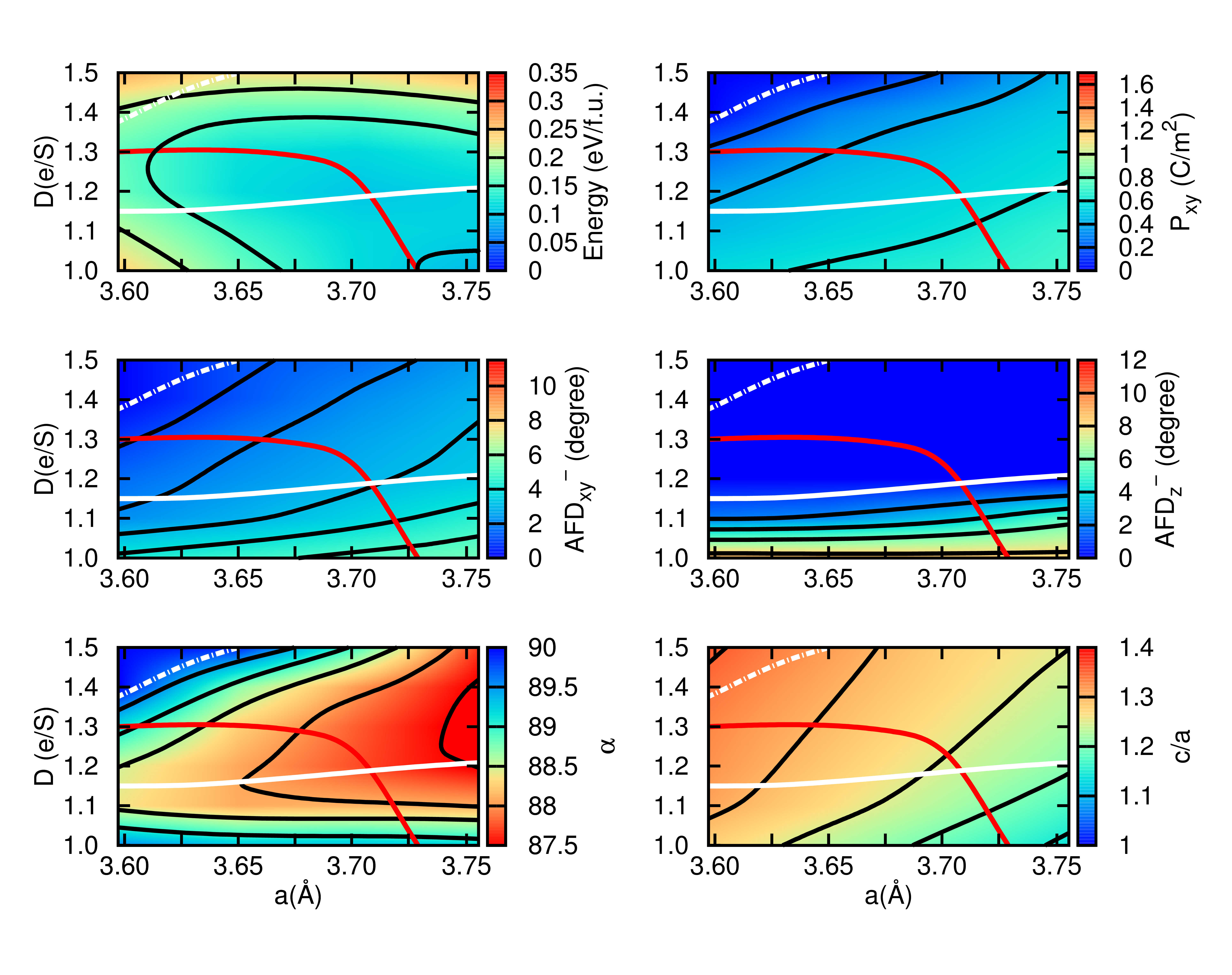}
\vspace{-0.75cm}
\caption{Same as in Fig.~\ref{figbfo-1}, but considering smaller (larger) $a$ ($D$)        
         values.}
\label{figbfo-2}
\end{figure}

For the case of BFO, we shall present our results by following a similar 
strategy to what we did in the PTO case: The overall phase diagram is shown in 
Fig.~\ref{bfo-pd}, while Figs.~\ref{figbfo-1} and~\ref{figbfo-2} illustrate the 
detailed behavior of the individual degrees of freedom. (We split the relevant 
parameter space into two parts here, to better describe the large compressive strain
region, where the spontaneous polarization adopts very large values.)
We use the same convention as in Figs.~\ref{pto-pd} and~\ref{figpto-1} regarding
the electrostatic equilibrium path and the two-phase boundaries, with the only
difference that, contrary to the PTO case, the latter are almost always shown as solid lines
here.
This reflects an important outcome of our calculations: most phase boundaries in 
BFO turn out to be of the first-order type. (The only exception is the transition
to the $P4mm$ state at very small values of $a$, marked in the upper left corner
of Fig.~\ref{bfo-pd}.)

As before, the path described by the $E_z=0$ states (minimum energy configurations at a
given value of $a$) is consistent with the results of earlier works~\cite{yang12,stengel12}: 
(i)~For a large range of in-plane strain values around the equilibrium parameter, BFO
adopts a monoclinic $Cc$ (we shall indicate this phase as $Cc$-I) state 
that is closely related to the bulk rhombohedral ground state. 
Here, both the polarization and the AFD$^-$ vector are 
large and roughly oriented along the (111) pseudocubic direction. 
Their evolution with decreasing $a$ follows the usual trend: $P_z$ and AFD$^-_z$ 
display a progressive enhancement, while the in-plane components of both
degrees of freedom simultaneously decrease in amplitude. 
(ii)~At a sufficiently large compressive strain ($a = 3.85$~\AA), the AFD$^-_z$ distortion 
amplitude peaks and then rapidly falls to zero, while both the $c/a$ ratio and $P_z$ 
undergo a drastic increase. (At $a = 3.65$~\AA, for instance, the value of $P_{z}$ 
is $1.56$~C/m$^{2}$ and $c/a = 1.28$~.) We identify this abrupt change with the 
isosymmetric~\cite{hatt10b} transition to the well-known \emph{supertetragonal} 
phase of BFO (also known in the literature as ``T-like'') that was experimentally observed 
in epitaxial films under a large compressive strain~\cite{bea09,zeches09}.
(iii)~At the other extreme end of the phase diagram (i.e. in the strongly tensile regime, $a > 4.07$~\AA)
BFO stabilizes in an orthorhombic $Pmc2_{1}$ phase, in all respects analogous to the
$Pmc2_{1}$ structure that occurs in PTO at similarly large strain values (see previous section). 

When we move away from the electrostatic equilibrium curve, BFO reveals a number
of remarkable (and potentially useful) features, as we shall illustrate in
the following.
In the zero or mildly compressive strain regime ($a \le 3.92$ \AA) 
we observe at small $D$ values a region where BFO adopts an orthorhombic
$Pna2_{1}$ structure. This state exhibits very large AFD$_{z}^{+}$ and 
AFD$_{xy}^{-}$ O$_{6}$ rotations, and a nonzero $P_z$. The latter, however,
is not ``ferroelectric'' in the usual sense: the system here behaves like a
nonlinear dielectric, with a metastable local minimum at $D=0$, where the
symmetry becomes orthorhombic $Pnma$. We can thus expect that under a strong
depolarizing field (and conditions that prevent domain formation) a BFO film 
might adopt a $Pnma$ ground state, which was so far only observed in bulk
at high pressure~\cite{haumont09}.
As $a$ is decreased, the $D$-interval over which this orthorhombic phase occurs 
becomes broader, while the amplitude of the AFD$_{z}^{+}$ and AFD$_{xy}^{-}$ oxygen 
rotations remain practically unchanged.
At even larger compressive strains ($3.60 \le a \le 3.65$~\AA) we find a new
region, this time at \emph{large} values of $D$, where
the AFD$_{xy}^{-}$ oxygen octahedral rotations and in-plane polarization completely 
disappear and the crystal becomes tetragonal $P4mm$. 
According to our calculations, the electric field-induced monoclinic 
$Cc$-II~$\to$~tetragonal $P4mm$ transformation is of second-order type. 
Note that this same $P4mm$ phase could be,
in principle, obtained even without the application of an electric field, by 
further compressing the crystal in-plane~[\onlinecite{stengel12}].
However, the strain that one would have to apply is enormous ($a \le 3.60$~\AA) --
our calculations show that, by \emph{combining} the effects of in-plane compression
and out-of-plane electric field, one might have better chances of experimentally
stabilizing the high-energy $P4mm$ phase, which is probably out of reach if
one relies solely on mechanical means.
Finally, moving now to the large tensile strain region ($a>4.07$ \AA), 
BFO adopts the same monoclinic $Pc$ phase that we have found in PTO under similar conditions 
(see Sec.~\ref{subsec:pto}). 
Also in this case it is interesting to note that the AFD$_{z}^{+}$ O$_{6}$ rotations become 
larger in amplitude as $D$, and consequently $P_{z}$, is increased. Remarkably, the
potential energy surface around $D=0$ appears to be rather flat, which indicates
a very large dielectric polarizability; this has important implications for practical
applications, as we shall discuss shortly in Sec.~\ref{subsec:funct}.

\subsection{Functional properties}
\label{subsec:funct}

\begin{figure}
\centerline
        {\includegraphics[width=1.0\linewidth]{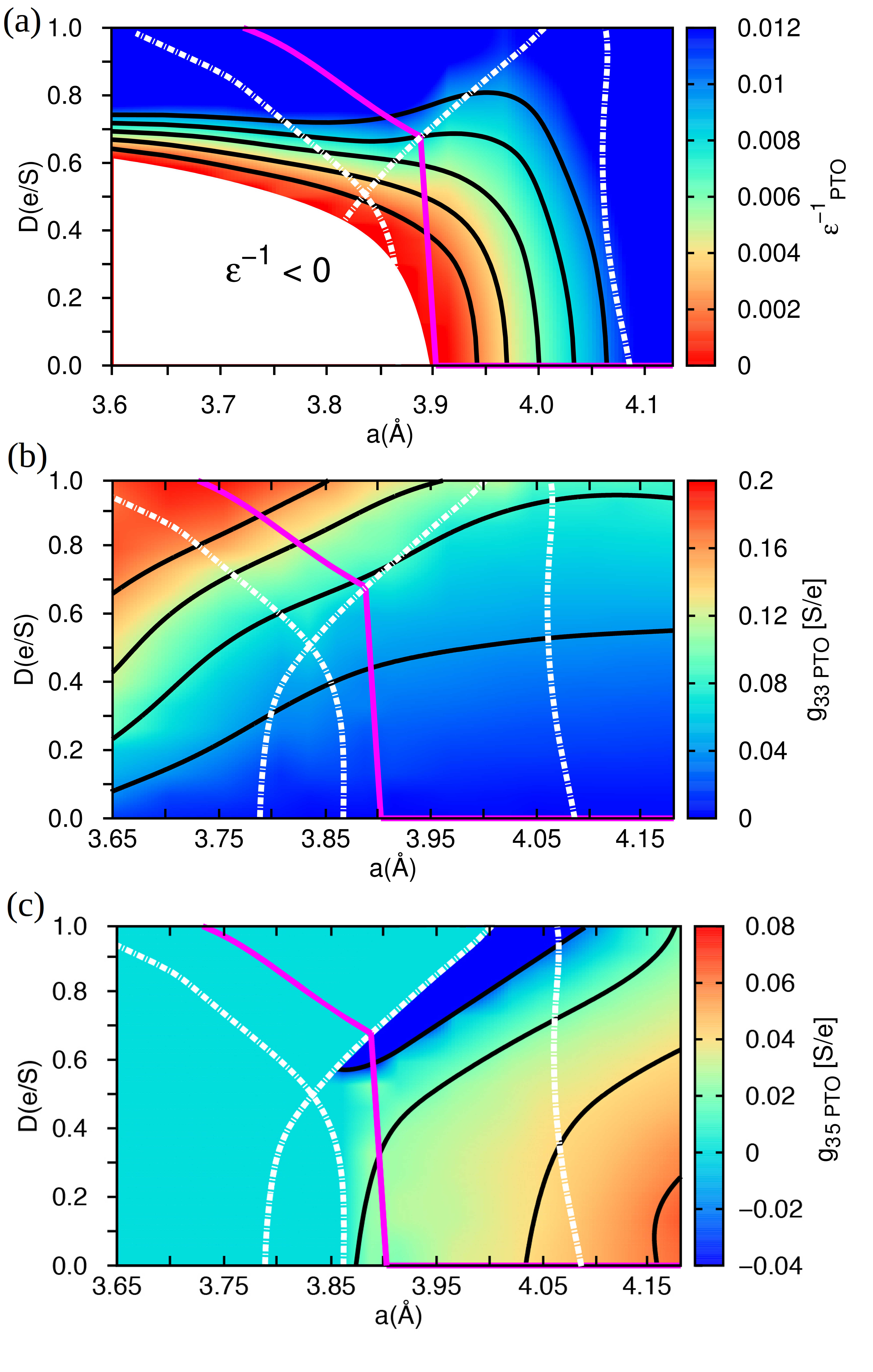}}
\vspace{-0.25cm}
\caption{Inverse dielectric constant $\epsilon^{-1}$~(a), and fixed-$D$ 
         piezoelectric coefficients $g_{33}$~(b) and $g_{35}$~(c), 
         calculated in PTO and expressed as a function of $a$ and $D$. The instability 
         region wherein the dielectric constant would be negative is not shown~(a). 
         The magenta solid line joins the states of minimum energy at given $a$,
         and the dashed white lines mark the boundaries between different phases.}
\label{pto-funct}
\end{figure}

The results described above have direct implications over the functional properties
of the two materials, as we shall see in the following. We shall focus more specifically
on the dielectric and piezoelectric properties, whose dependence on the external  parameters 
will be again illustrated in the form of 2D maps. We shall discuss the two materials 
separately, starting with PbTiO$_3$.

\subsubsection{Bulk PbTiO$_{3}$}
\label{subsec:ptofunc}

In Fig.~\ref{pto-funct}a, we show the calculated inverse dielectric constant, $\epsilon^{-1}$, 
of bulk PTO in the region of parameter space spanned by 
$3.6 \le a \le 4.2$~\AA~ and $0 \le D \le 1.0~e/S$. 
First of all, note the blank area in the lower left corner of the graph;
it corresponds to $(a,D)$ points where $\epsilon^{-1}$ adopts negative values,
pointing to an instability of the system towards a ferroelectric distortion.
As these points are experimentally inaccessible, we have decided to exclude 
them from our analysis.
The most interesting region in the $(a, D)$ map is that where the 
crystal, under zero-field conditions, transitions (for increasing $a$) from
the orthorhombic $Amm2$ to the tetragonal $P4mm$ transition via 
the intermediate monoclinic $Cm$ phase (see Sec.~\ref{subsec:pto}). 
Here 
In the $3.90 \le a \le 3.94$~\AA~ 
interval, for instance, $\epsilon$ adopts values which are higher than $\sim 500$ (in units of 
$\epsilon_{0}$, that is, the dielectric constant of vacuum) in the corresponding zero-field 
states (indicated with a magenta line). 
The $Amm2$~$\to$~$P4mm$ phase transition corresponds to an abrupt rotation of
the polarization vector from in-plane to out-of-plane, and thus the large dielectric 
response of the system in that region stems from the flatness of the potential landscape
that governs $P_z$. 
Indeed, in the tensile-strain regime the system behaves as a normal dielectric along $z$,
i.e. the internal energy of the system $U(D)$ depends quadratically on $D$. Conversely, in the 
compressive regime, $U(D)$ adopts a symmetric double-well form (see Fig.~\ref{pto-cm}). 
In crossing the boundary, $U(D)$ flattens significantly in the region near $D = 0$, where its 
second derivative, and consequently $\epsilon^{-1}$, becomes very small. 

As they are directly proportional to $\epsilon$, the $d_{33}$ and $d_{35}$ piezoelectric coefficients 
of bulk PTO are expected to be giant (see Eqs.~\ref{eq:d33} and~\ref{eq:d15} in 
Sec.~\ref{subsec:funct-prop}) in the transition region described above. Of course, a necessary 
condition is that the corresponding $g_{33}$ and $g_{35}$ coefficients do not vanish.
Regarding the longitudinal $g_{33}$ coefficient, Fig.~\ref{pto-funct}b shows a relatively simple
monotonous behavior, with a roughly linear increase from zero (at the $D=0$ state, where it vanishes 
by symmetry) to $0.08-0.20$ (depending on the strain state) at the highest values of $D$. 
(Interestingly, the rate of increase of $g_{33}$ with $D$, which can be regarded as
an effective electrostrictive coefficient, seems to be rather sensitive to 
the in-plane strain, being largest in the extreme compressive regime, and milder
under tensile conditions.)
By combining the information of $\epsilon^{-1}$ and $g_{33}$, one can draw 
the following conclusions: (i) $d_{33}$ vanishes at tensile strain by symmetry,
unless a sufficiently large electric field is applied to the crystal. (ii)
$d_{33}$ is nonzero but relatively small in the compressive strain regime,
where the zero-field curve lies far from the region where $\epsilon$ 
diverges. (One can, in principle, try to approach that region by applying 
an $E$-field antiparallel to $P_z$, although the crystal will likely
switch to the opposite polarization state well before reaching it.) (iii)
In the intermediate region where PTO is either monoclinic or tetragonal
(but close to the phase boundary), $d_{33}$ becomes giant as speculated above,
due to the structural and dielectric softness of the system therein.

As for the \emph{shear} piezoelectric coefficient, $g_{35}$ (see Fig.~\ref{pto-funct}c), we find
a very rich behaviour that is remarkably sensitive to the mechanical and electrical boundary conditions.
First, note that whenever $P_{xy}$ vanishes (i.e. at compressive strain), 
$g_{35}$ vanishes because of symmetry (the crystal axes form orthogonal angles
with no shear). 
The physically interesting region occurs from zero to moderately tensile strain,
where the in-plane components of the polarization start being active.
Surprisingly, the $g_{35}$ coefficient shows a rather peculiar behavior: it is positive
in most of the region of parameter space where $P_{xy}$ is nonzero (this includes all 
equilibrium zero-field states with $Amm2$ symmetry, occurring at $a>3.90$~\AA), except for 
a narrow region close to the phase boundary where it becomes \emph{negative}.

In order to understand the physical origin of this result, it is useful to recall
the form of the coupling terms in the free energy that involve polarization and shear strain.
At the lowest order, in PTO such a coupling is governed by~\cite{pertsev98}
\begin{equation}
E_{\rm coupl} = A P_z P_{xy} \sigma_{\rm shear},
\end{equation}
where $\sigma_{\rm shear}$ is the corresponding component of the
stress tensor (note that the shear occurs between the out-of-plane [001] and 
in-plane [110] axes), and $A$ is the coupling coefficient. The $g_{35}$ 
piezoelectric coefficient can be roughly estimated by deriving the above 
coupling term twice with respect to $P_z$ (which in the context of our
calculations essentially corresponds to $D$) and $\sigma_{\rm shear}$,
\begin{equation}
g_{35} \sim \frac{ \partial^2 E_{\rm coupl}}{\partial P_z \partial \sigma_{\rm shear}}.
\end{equation}
It is useful, for analysis purposes, to break down the above equation into two 
intermediate steps, by introducing the shear strain, $\varepsilon_{\rm shear}$,
\begin{equation}
\varepsilon_{\rm shear} \sim \frac{ \partial E_{\rm coupl}}{\partial \sigma_{\rm shear}}, \qquad 
g_{35} \sim \frac{ \partial \varepsilon_{\rm shear}}{\partial P_z}.
\end{equation}
One has, trivially,
\begin{equation}
\varepsilon_{\rm shear} \propto P_z P_{xy}.
\end{equation}
This behavior of $\varepsilon_{\rm shear}$, which corresponds to the deviation
of the monoclinic angle $\alpha$ from 90$^\circ$, is qualitatively well reproduced 
in our calculations (see Fig.\ref{figpto-1}).
From the evolution of $\alpha$ as a function of $D$ (for a fixed $a$), one can 
immediately appreciate the origin of the sign change in $g_{35}$: close to $D=0$, $\alpha$ grows 
linearly with $D$ (i.e. its derivative with respect to $D$ is positive), reaches a 
peak and then rapidly drops back (negative region) to zero when approaching
the phase boundary.

In a close vicinity of such boundary, $P_z$ has a finite value, and therefore  
can be approximated by a constant; one should then have 
\begin{equation}
\varepsilon_{\rm shear} \propto P_{xy}.
\label{eq1}
\end{equation}
Now, since the transition is of second-order type, for a given $a$ one expects a 
critical behavior of $P_{xy}$ with respect to $D$ of the form
\begin{equation}
P_{xy} \propto \sqrt{D_0(a) - D},
\label{eq2}
\end{equation}
where $D_0(a)$ describes the phase boundary. By combining Eq.~(\ref{eq1}) and 
Eq.~(\ref{eq2}), and by further taking the first derivative in $D$, one obtains
that the piezoelectric coefficient $g_{35}$ \emph{diverges} at the boundary as
\begin{equation}
g_{35} \propto -\frac{1}{\sqrt{D_0(a) - D}}.
\end{equation}

Unfortunately, the divergence does not show up in our graphs because of the relatively
coarse grid of $(a,D)$ points that we have used to sample our parameter
space.
Nevertheless, the above conclusion is solid, as it stems from basic
symmetry arguments and from the second-order nature of the phase transition.
(The latter is unambiguously demonstrated by our numerical results.)
This points to a promising route to achieving a large piezoelectric response
by tuning the $D$-coefficient $g_{35}$, rather than the dielectric constant.
We expect this mechanism to be especially useful in superlattices,
because of the additive property [see Eq.~(\ref{eq:gadd})] of the $g$-coefficients.
[Conversely, the dielectric constant obeys the typical series-capacitor
model, Eq.~(\ref{eq:gadd}), making it harder to enhance this latter 
functionality in a layered system.]

\subsubsection{Bulk BiFeO$_{3}$}
\label{subsec:bfofunc}

The calculated inverse dielectric constant of BFO, expressed as a function of 
the in-plane lattice parameter and out-of-plane electric displacement, is shown 
in Fig.~\ref{bfo-funct}a. 
As in the PTO case, we have excluded from the graph the region where the system is unstable against
a polar distortion, and hence has a negative permittivity (white area). (In the region where BFO 
adopts the $Pna2_1$ phase $\epsilon^{-1}$ is in fact positive however, for the sake of simplicity 
and because that region is experimentally difficult to access, we ignore it in the 
following analysis). The data points that lie closest to the white area are characterized by a vanishing 
value of $\epsilon^{-1}$, and hence by a diverging dielectric constant, which could be interesting for 
applications. 
At large tensile strain, $\epsilon^{-1}$ presents some discontinuities,
which occur more specifically along the $Pmc2_{1}$~$\to$~$Cc$-I phase boundary. 
(Recall that the involved phase transformation is of first-order type, and
is characterized by a drastic structural change; this explains the abrupt jump
in the response properties of the crystal.)
Interestingly, at $a > 4.05$~\AA and for moderate values of $D$ ($0 \le D \le 0.5$~$e/S$), 
where the system is in a monoclinic $Pc$ phase (orthorhombic $Pmc2_{1}$ at $D=0$),
$\epsilon$ adopts very large values. 
In fact, the high-tensile strain $Pmc2_{1}$ phase appears to be on the
verge of having a polar instability along $z$. This fact reflects itself
in an unusual flatness of the $U(D)$ energy curves around $D=0$, which in turn
results into a very large dielectric (and piezoelectric, as we shall see shortly) 
response of the system (see Eq.~\ref{eq:dielectric}). 

\begin{figure}
\centerline
        {\includegraphics[width=1.0\linewidth]{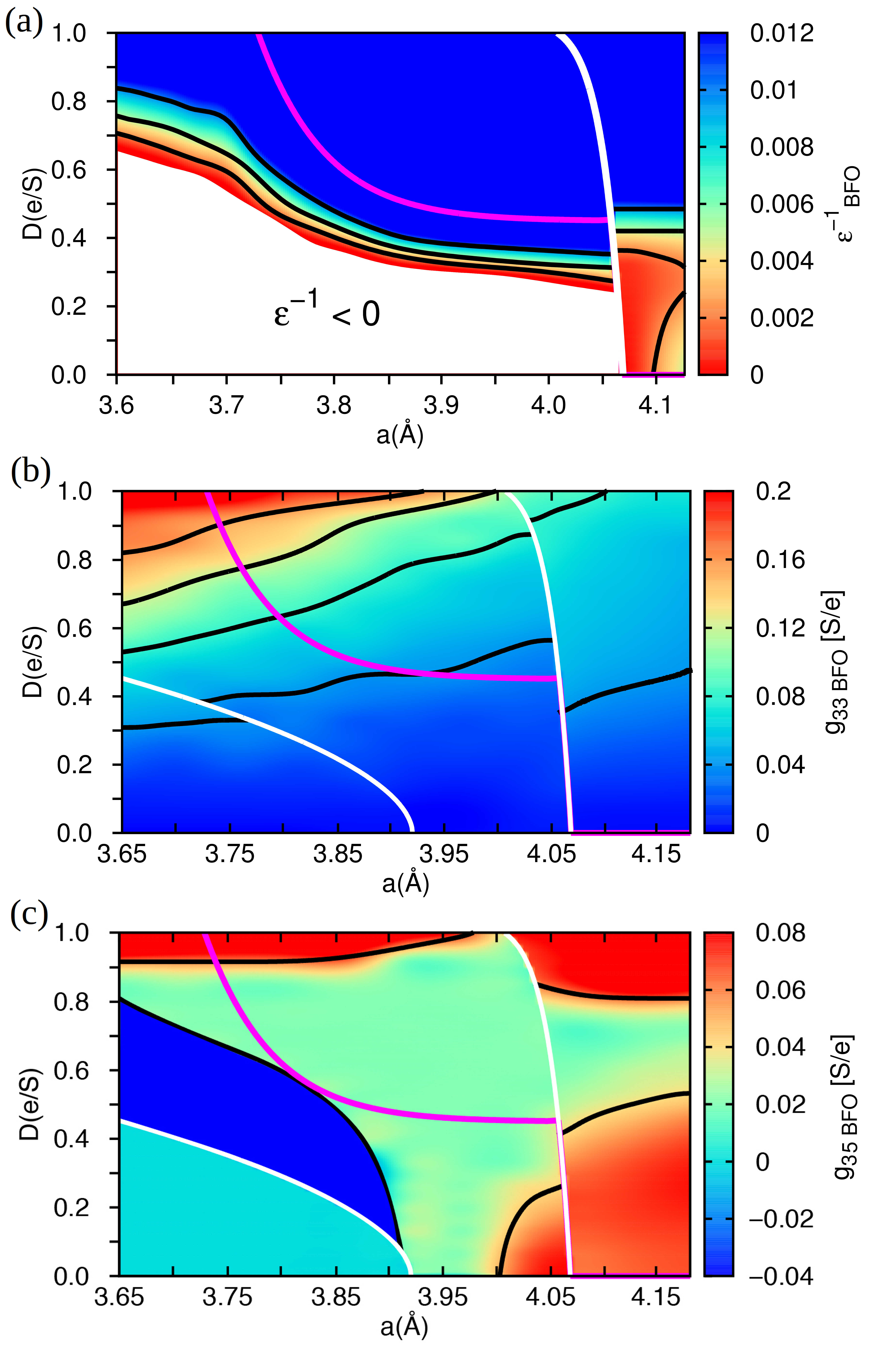}}
\vspace{-0.25cm}
\caption{Inverse dielectric constant $\epsilon^{-1}$~(a) and the fixed-$D$   
         piezoelectric coefficients $g_{33}$~(b) and $g_{35}$~(c) calculated 
         in BFO and expressed as a function of $a$ and $D$. The instability
         region wherein the dielectric constant would be negative is not shown~(a). 
         The magenta solid line joins the states of minimum energy at given $a$,
         and the solid white lines mark the boundaries between different phases.}
\label{bfo-funct}
\end{figure}

Regarding the \emph{longitudinal} piezoelectric coefficient (see Fig.~\ref{bfo-funct}b), 
$g_{33}$, BFO behaves roughly like PTO: $g_{33}$ is zero at $D=0$, grows linearly
for increasing $D$ and is largest in the upper left corner of the plot, i.e.
at large compressive strain and large out-of-plane polarization.
(Of course, the first-order phase transition lines and the corresponding
discontinuities in $g_{33}$ are absent in PTO; however, these do not 
produce a profound qualitative change in the overall picture.)
Note that, at difference with PTO, the system does not cross a 
state of marked structural softness at small applied strains 
(the zero-field line stays far away from the high-$\epsilon$
region unless a very large tensile strain is applied, while in
PTO it crosses through it at $a \sim 3.9$~\AA, see Fig.~\ref{pto-funct}a).
Thus, as in the large tensile strain region $g_{33}$ (and hence $d_{33}$)
vanishes by symmetry at zero electric field, we conclude that the 
transition from $Cc$-I to $Cc$-II (which occurs at large compressive strain)
constitutes the most promising configuration for obtaining an enhanced value of
$d_{33}$ in BFO.

The \emph{shear} piezoelectric coefficient, on the other hand, displays a complex 
behavior that differs qualitatively from the PTO case and that does not lend 
itself to a comparatively simple physical interpretation (see Fig.~\ref{bfo-funct}c).
As a matter of fact in BFO we have, in addition to the polarization-strain coupling
that we mentioned in the previous Section, a further term in the free energy that
couples the strain to the AFD modes,
\begin{equation}
E_{\rm coupl} = A P_z P_{xy} \sigma_{\rm shear} + B \phi_z \phi_{xy} \sigma_{\rm shear},
\end{equation}
where $B$ is a second coupling coefficient and $\phi_{z,xy}$ are the AFD$^{-}_{z,xy}$ 
distortion amplitudes.
It is difficult, in the context of our calculations, to disentangle the effects
of the two different coupling terms; for this reason, in the following we shall 
limit ourselves to commenting on the general trends that emerge from our data.

To start with, we note that the system in the proximity of zero epitaxial strain 
is characterized by small $g_{35}$ values (the zero-field curve stay close
to the $g_{35}=0$ isoline for a relatively wide range of in-plane lattice 
parameters.
Again, only at rather extreme values of the epitaxial strain does the 
system become potentially interesting for piezoelectric applications.
For example, at high compressive strains the shear angle shows a steep increase as 
a function of $D$ (see Fig.~\ref{figbfo-2}), which results in a relatively large 
$g_{35}$ coefficient.
(Recall that in similar conditions the out-of-plane $c$ parameter is also characterized
by a strong dependence on $D$ -- this yields a large value of the longitudinal $g_{33}$
coefficient, as we pointed out earlier.)
The most interesting region of parameter space in the present context, however, is by far the
tensile strain one, where BFO adopts the $Pmc2_1$ phase.
Here a relatively large value of $g_{35}$ is combined with a giant value of the 
out-of-plane dielectric constant, $\epsilon$; this results in a $d_{35}$
coefficient in excess of $700$~pm/V at zero applied field.

\section{Summary}
\label{sec:summary}
We have studied the physical behaviour of archetypal ferroelectrics PbTiO$_{3}$ 
and BiFeO$_{3}$ under arbitrary electrical and mechanical boundary conditions, by means of 
state-of-the-art \emph{ab initio} methods. 
Our calculations provide a 
detailed description of the complex interplay between polar and structural
degrees of freedom in these materials over a comprehensive range of in-plane strain
$(a)$ and out-of-plane electric displacement field $(D)$ values. 
Of particular relevance is the assessment and rationalisation of the related dielectric 
and piezoelectric response functions, which can be naturally accessed as a by-product of our 
calculations.  

In bulk PTO, we have predicted the existence of a new monoclinic $Cm$ phase that is stable 
in a narrow strain interval, and that bears important similarities to a previously reported 
structure in PTO-based superlattices. 
We have identified the regions in $(a, D)$ space wherein PTO exhibits a giant dielectric and 
piezoelectric response. In particular, we have unravelled a 
coupling mechanism by which its shear piezoelectric coefficient $d_{35}$ diverges in reaching 
the monoclinic $Cm$--tetragonal $P4mm$ phase boundary. 
In BFO, we have shown that a complex 
interplay between polar and antipolar distortions reigns in the region of high tensile strains 
wherein an orthorhombic $Pmc2_{1}$ phase becomes stable. In that same region, we find that 
both the dielectric and shear piezoelectric responses of the system are extremely large.

While the present analysis has been based on bulk calculations, we emphasise 
that our findings are directly relevant to the the rational 
design of multicomponent ferroelectric superlattices, as our methodology
allows one to accurately predict their functional properties based on the
$(a,D)$ maps of the bulk building blocks. 
In this sense, our study opens a promising new avenue in the ongoing
search for new functional oxide systems with enhanced functionalities.

\acknowledgments
This research was supported by the Australian Research Council's
Future Fellowship funding scheme (project numbers RG134363 and RG151175),
MINECO-Spain (Grants No. MAT2010-18113, No. CSD2007-00041, and No. 
FIS2013-48668-C2-2-P), and Generalitat de Catalunya (2014 SGR301).
We thankfully acknowledge the computer resources, technical expertise and 
assistance provided by RES, CESGA, and the National Computational Infrastructure
supported by the Australian Government.


\begin{thebibliography}{30}
\bibitem{cohen90} R. E. Cohen and H. Krakahuer, Phys. Rev. B \textbf{42}, 6416 (1990). 
\bibitem{cohen92} R. E. Cohen, Nature \textbf{358}, 136 (1992).
\bibitem{junquera12} P. Aguado-Puente, P. Garc\'ia-Fern\'andez, and J. Junquera,
                     Phys. Rev. Lett. \textbf{107}, 217601 (2011).
\bibitem{bousquet08} E. Bousquet, M. Dawber, N. Stucki, C. Lichtensteiger, P. Hermet,
                     S. Gariglio, J.-M. Triscone, and P. Ghosez, Nature (London)
                     \textbf{452}, 732 (2008).
\bibitem{levanyuk74} A. P. Levanyuk and D. G. Sannikov, Sov. Phys. Uspekhi, 
                     \textbf{17}, 199 (1974).
\bibitem{benedek11} N. A. Benedek and C. J. Fennie, Phys. Rev. Lett. \textbf{106}, 107204 (2011) 
\bibitem{benedek12} N. A. Benedek, A. T. Mulder, and C. J. Fennie, 
                    J. Sol. Stat. Chem. \textbf{195}, 11 (2012). 
\bibitem{heron14} J. T. Heron \emph{et al.}, Nature \textbf{516}, 370 (2014).
\bibitem{fusil14} S. Fusil, V. Garc\'{i}a, A. Barth\'{e}l\'{e}my, and M. Bibes,
                  Annu. Rev. Mater. Res. \textbf{44}, 91 (2014). 
\bibitem{hatt10} A. J. Hatt and N. A. Spaldin, Phys. Rev. B \textbf{82}, 195402 (2010).
\bibitem{rondinelli11} J. M. Rondinelli and N. A. Spaldin, 
                       Adv. Mater. \textbf{23}, 3363 (2011). 
\bibitem{rondinelli11b} J. M. Rondinelli and S. Coh, 
                       Phys. Rev. Lett. \textbf{106}, 235502(2011).
\bibitem{may10} S. J. May, J. -W. Kim, J. M. Rondinelli, E. Karapetrova, N. A. 
                Spaldin, A. Bhattacharya, and P. J. Ryan, Phys. Rev. B
                \textbf{82}, 014110 (2010).
\bibitem{wu12} C. W. Swartz and X. Wu, Phys. Rev. B \textbf{85}, 054102 (2012).
\bibitem{stengel12b} M. Stengel, C. J. Fennie, Ph. Ghosez, 
                     Phys. Rev. B \textbf{86}, 094112 (2012).
\bibitem{hong13} J. Hong and D. Vanderbilt, Phys. Rev. B \textbf{87}, 064104 (2013).
\bibitem{dawber05b} M. Dawber, K. M. Rabe, and J. F. Scott, 
                    Rev. Mod. Phys. \textbf{77}, 1083 (2005).
\bibitem{choi04} K. J. Choi \textit{et al.}, Science  \textbf{306}, 5698 (2004).
\bibitem{bea09} H. B${\rm \acute{e}}$a, B. Dup${\rm \acute{e}}$, S. Fusil, R. Mattana, E. Jacquet,
                B. Warot-Fonrose, F. Wilhelm, A. Rogalev, S. Petit, V. Cros, A. Anane,
                F. Petroff, K. Bouzehouane, G. Geneste, B. Dkhil, S. Lisenkov, I. Ponomareva,
                L. Bellaiche, M. Bibes, and A. Barth${\rm \acute{e}}$l${\rm \acute{e}}$my,
                Phys. Rev. Lett. \textbf{102}, 217603 (2009).
\bibitem{warusawithana09} M. P. Warusawithana \emph{et al.}, Science \textbf{324}, 367 (2009).
\bibitem{jacek10} J. C. Wojde\l and J. ${\rm \acute{I}}$${\rm \tilde{n}}$iguez,
                    Phys. Rev. Lett. \textbf{105}, 037208 (2010).
\bibitem{junquera11} C. Lichtensteiger \textit{et al.}, in \textit{Oxides Ultrathin Films: Science
                     and Technology}, edited by G. Pacchioni and S. Valeri,
                     Ch. 12, 265 (Wiley-VCH, Germany, 2011).
\bibitem{ghosez08} P. Ghosez and J. Junquera, J. Comp. Theor. Nanosci. \textbf{5}, 2071 (2008).
\bibitem{zubko12} P. Zubko \textit{et al.}, Nano Letters \textbf{12}, 2846 (2012).
\bibitem{dawber12} J. Sinsheimer \textit{et al.}, Phys. Rev. Lett. \textbf{109}, 167601 (2012).
\bibitem{dawber05} M. Dawber \textit{et al.}, Phys. Rev. Lett. \textbf{95}, 177601 (2005).
\bibitem{dawber07} M. Dawber \textit{et al.}, Adv. Mater. \textbf{19}, 4153 (2007).
\bibitem{souza02} I. Souza, J.  ${\rm \acute{I}}$${\rm \tilde{n}}$iguez, and D. Vanderbilt, 
                  Phys. Rev. Lett. \textbf{89}, 117602 (2002).
\bibitem{dieguez06} O.  Di\'eguez and D. Vanderbilt, 
                    Phys. Rev. Lett. \textbf{96}, 056401 (2006).
\bibitem{stengel09c} M. Stengel, N. A. Spaldin, and D. Vanderbilt, 
                     Nature Physics \textbf{5}, 304 (2009).
\bibitem{wojdel13} J. C. Wojde\l , P. Hermet, M. P. Ljungberg, P. Ghosez, and 
		   J. ${\rm \acute{I}}$${\rm \tilde{n}}$iguez, 
                   J. Phys.:Condens. Matter \textbf{25}, 305401 (2013). 
                    Phys. Rev. Lett. \textbf{105}, 037208 (2010).
\bibitem{yang12} Y. Yang, W. Ren, M. Stengel, X. H. Yan, and L. Bellaiche, 
                 Phys. Rev. Lett. \textbf{109}, 057602 (2012).
\bibitem{dieguez11} O. Di\'{e}guez, O. E. Gonz\'{a}lez-V\'{a}zques, J. C. Wojde\l, 
                    and J. ${\rm \acute{I}}$${\rm \tilde{n}}$iguez,
                    Phys. Rev. B \textbf{83}, 094105 (2011).
\bibitem{guennou11} M. Guennou, P. Bouvier, G. S. Chen, B. Dkhil, R. Haumont, G. Garbarino,
                    and J. Kreisel, Phys. Rev. B {\bf 84}, 174107 (2011).
\bibitem{prosandeev12} S. Prosandeev, D. Wang, W. Ren, J. ${\rm \acute{I}}$${\rm \tilde{n}}$iguez,
                       and L. Bellaiche, Adv. Funct. Mater. \textbf{23}, 234 (2013).
\bibitem{rahmedov12} D. Rahmedov, D. Wang, J. ${\rm \acute{I}}$${\rm
  \tilde{n}}$iguez, and L. Bellaiche, Phys. Rev. Lett. \textbf{109},
  037207 (2012).
\bibitem{cazorla13} C. Cazorla and J. ${\rm \acute{I}}$${\rm\tilde{n}}$iguez, 
                    Phys. Rev. B \textbf{88}, 214430 (2013).
\bibitem{zeches09} R. J. Zeches \textit{et al.}, Science \textbf{326}, 977 (2009).
\bibitem{stengel12} Y. Yang, M. Stengel, W. Ren, X. H. Yan, and L. Bellaiche, 
                    Phys. Rev. B \textbf{86}, 144114 (2012).
\bibitem{cazorla14} C. Cazorla and M. Stengel, Phys. Rev. B \textbf{90}, 020101(R) (2014).
\bibitem{kornev07} I. A. Kornev, S. Lisenkov, R. Haumont, B. Dkhil, and L. Bellaiche,
                   Phys. Rev. Lett. \textbf{99}, 227602 (2007).
\bibitem{stengel09} M. Stengel, D. Vanderbilt, and N. A. Spaldin,
                    Phys. Rev. B \textbf{80}, 224110 (2009).
\bibitem{wu05} X. Wu, D. Vanderbilt, and D. R. Hamann, Phys. Rev. B \textbf{72}, 035105 (2005).
\bibitem{dieguez04} O. Di\'{e}guez, S. Tinte, A. Antons, C. Bungaro, J. B. Neaton, K. M. Rabe, 
                    and D. Vanderbilt, Phys. Rev. B \textbf{69}, 212101 (2004).
\bibitem{blok11} J. L. Blok, D. H. A. Blank, G. Rijnders, K. Rabe and D. Vanderbilt, 
                 Phys. Rev. B \textbf{84}, 205413 (2011).
\bibitem{hatt10b} A. J. Hatt, N. A. Spaldin, and C. Ederer, 
                  Phys. Rev. B \textbf{81}, 054109 (2010).
\bibitem{haumont09} R. Haumont, P. Bouvier, A. Pashkin, K. Rabia, S. Frank, B. Dkhil, 
                    W. A. Crichton, C. A. Kuntscher, and J. Kreisel, 
                    Phys. Rev. B \textbf{79}, 184110 (2009).
\bibitem{pertsev98} N. A. Pertsev, A. G. Zembilgotov, and A. K. Tagantsev, Phys. Rev. Lett. \textbf{80}, 1988 (1998). 
\end{thebibliography}
\end{document}